\documentstyle[12pt]{article}

\newcommand{\nc}{\newcommand}
\def\frac#1#2{{\textstyle {#1 \over #2}}}

\nc{\beq}{\begin{equation}}
\nc{\eeq}{\end{equation}}
\nc{\beqa}{\begin{eqnarray}}
\nc{\eeqa}{\end{eqnarray}}
\nc{\lsim}{\begin{array}{c}\,\sim\vspace{-21pt}\\< \end{array}}
\nc{\gsim}{\begin{array}{c}\sim\vspace{-21pt}\\> \end{array}}

\def\&{and}

\def\CR{\nonumber \\ }

\def\Dslash{\not{\hbox{\kern-4pt $D$}}}

\def\L{{\cal L}}

\def\O{{\cal O}}
\def\W{{\cal W}}

\def\ie{{\it i.e. }}
\def\eg{{\it e.g. }}

\def\nc#1#2#3{           {\it Nuovo Cim.  }{\bf #1}, #2 (19#3)}
\def\np#1#2#3{           {\it Nucl. Phys. }{\bf #1}, #2 (19#3)}
\def\pl#1#2#3{           {\it Phys. Lett. }{\bf #1}, #2 (19#3)}
\def\pr#1#2#3{           {\it Phys. Rev. }{\bf #1}, #2 (19#3)}
\def\prep#1#2#3{         {\it Phys. Rep. }{\bf #1}, #2 (19#3)}
\def\prl#1#2#3{          {\it Phys. Rev. Lett. }{\bf #1}, #2 (19#3)}

\begin{document}

\title{\large{\bf
A Note on Supersymmetry Breaking}}

\author{
Stephen D.H.~Hsu\thanks{hsu@hsunext.physics.yale.edu}, \\
Myckola Schwetz\thanks{ms@genesis2.physics.yale.edu}
\\ \\ Department of Physics \\ Yale University
\\ New Haven, CT 06520-8120 }


\date{March, 1997}

\maketitle

\begin{picture}(0,0)(0,0)
\put(350,315){YCTP-P3-97}
\end{picture}
\vspace{-24pt}

\begin{abstract}
Using a simple observation based on holomorphy, we argue that any model
which spontaneously breaks supersymmetry for some range of a parameter
will do so generically for all values of that parameter, 
modulo some isolated exceptional
points. Conversely, a model which preserves supersymmetry for some range
of a parameter will also do so everywhere except at isolated exceptional points.
We discuss how these observations can be useful in 
the construction of new models 
which break supersymmetry and discuss some simple examples. We also
comment on the relation of these results to the Witten index.

\end{abstract}

\newpage

\section{Holomorphy and SUSY breaking}

Recently, due to renewed interest and improved techniques \cite{Seiberg},
there has been a great deal of progress in our understanding of spontaneous
(dynamical) supersymmetry breaking. The number of models which 
are thought to break
SUSY has increased substantially \cite{DNS,ISS,IY,IT,models} relative
what was known a decade ago \cite{ADS}.
In this note 
we wish to examine the question of SUSY breaking in the parameter space
of a specified model with fixed field content and interactions. We will argue that
the generic behavior in the whole of parameter space can be deduced from
that of any small patch. That is, models either do or do not break SUSY in
the entire parameter space, with the possible exception of isolated
points in the space which form a set of measure zero. This conclusion is
similar to that obtained from the Witten index \cite{WI} in the case of 
$\rm{Tr(-1)^F} \neq 0$ --
$\ie$ when SUSY is unbroken. However, the picture we obtain extends also
to models in which $\rm{Tr(-1)^F} = 0$ and the status of SUSY breaking is still
ambiguous.

The argument for this is simple and based on holomorphy. Consider the 
Wilsonian effective action for a model which possesses $N=1$ SUSY:
\beqa
\label{Lag}
\L_{\rm eff} &=& \int d^4\theta \,K(\Phi^\dagger e^{V}\Phi) ~+~
\frac{1}{8\pi} Im \int
d^2\theta\,\tau (\Phi)\, \W^{\alpha}\W_{\alpha} \CR
& &+~2 Re \int
d^2\theta\, W(\Phi) ~+~ \{ {\rm higher ~dim. ~\W^2} \}~,
\eeqa
with K\"{a}hler metric
\beq
\label{KM}
g_{ij} = { {\partial^2 K} \over {\partial \Phi^{\dagger i}\partial\Phi^j} }\ .
\eeq
In the last term in (\ref{Lag}) we include higher dimension $D$- and $F$-terms 
which depend on $\W^{\alpha}$  ($K$ and $W$ already 
include higher dimension operators
in $\Phi_i$). For simplicity below we will focus on $\Phi_i$ in our discussion, but
our conclusions will apply also to the chiral superfield
$\W^{\alpha} \W_{\alpha}$, and hence to
gaugino condensation.

It is important to note that a supersymmetric 
effective Lagrangian description of the
form (\ref{Lag}) will apply whether or not SUSY is spontaneously broken.
As in the usual case of a spontaneously broken symmetry, the
interactions themselves respect the symmetry, and it is only 
the ground state that breaks the symmetry.
The relevant energy scale here is the scale at which SUSY is broken, 
so the superfield whose $F$-component 
acquires a vacuum expectation value will still appear in (\ref{Lag}).

%

Now consider the vacuum expectation value of the $F$-component 
of one of the chiral superfields:
\beq
\label{Fvev}
\langle F_j^* \rangle ~=~  ( g_{ij} )^{-1}~ { \partial W \over \partial \Phi_i }~
=~ ( g_{ij} )^{-1} W_{\Phi_i}.
\eeq
(A similar equation holds for gauge-invariant chiral superfields 
formed by products of the $\Phi_i$. It is perhaps more correct to use
such fields as order parameters for SUSY breaking.) 
Supersymmetry will be broken if any
of the $\langle F_i^* \rangle$ are non-zero. Suppose that this is established 
for some range of parameters (bare couping $\tau_0$, masses $m_i$, etc.). This
implies that the function $W_ {\Phi_i }$ is non-zero in that
range, and since
the superpotential $W$ depends holomorphically on these parameters, 
$W_{\Phi_i }$
cannot vanish outside the range 
except at some isolated points where it crosses through zero.
However, modulo singularities in the K\"{a}hler metric, this implies that supersymmetry
can only be restored at some isolated points in parameter space, and will
generically remain broken as the parameters of the model are varied.
(Here and below, by ``point'' we mean some subspace of lower dimensionality
than the entire parameter space. Along the relevant direction
orthogonal to this subspace the exceptional
behavior will only occur at a point.)

We can also apply this line of reasoning to models which preserve supersymmtery
in some patch of parameter space. In that case all $\langle F_i^* \rangle$
vanish in the patch, and hence
everywhere by holomorphy. Exceptions again come
from singularities in $g_{ij}$.

It is possible that the K\"{a}hler metric can become singular at certain
points, for example when new exactly massless modes 
appear in the low energy theory.
However, we expect that this again only occurs at an isolated point, 
and not in an entire patch of parameter space. 
Consider approaching the point 
$g_*$ in parameter space where a new massless
excitation (or multiplet of excitations) 
enters the low-energy theory. Near this special point, we should be
able to write a new effective Lagrangian with the extra particle included.
This new effective Lagrangian should have non-singular K\"{a}hler metric, and
the mass of the new excitation should appear as a term in the superpotential.
Now consider varying the relevant parameter past the point $g_*$. Since 
the particle mass is non-zero on one side of $g_*$, it cannot remain zero
on the other side due to holomorphy. Hence, away from $g_*$ there is no
exactly massless excitation and therefore the original K\"{a}hler metric was only
singular at $g_*$ and not in an entire neighborhood. This is essentially what
happens in $N=2$ SQCD at the singularities in the complex $u$ plane \cite{SW1}.
Of course, this discussion is in no way conclusive -- we can not rule out
the possibility of more pathological behavior of the K\"{a}hler metric.
However,
most analyses of SUSY dynamics concentrate on the superpotential, and
assume that the K\"{a}hler metric is well-behaved. Under those assumptions
one always finds
the type of behavior that we emphasize in this note.

These conclusions are very similar to Seiberg's
\cite{SW2} observation that phase transitions are forbidden by holomorphy
in SUSY models (again, modulo singularities in the K\"{a}hler metric). 
Seiberg's argument focuses on the ground state energy
\beq
\label{pot}
V ~=~ \sum_{ij} ~( g_{ji} )^{-1} ~ {\partial W \over \partial \Phi_i }
{\partial W^* \over \partial \Phi^{\dagger}_j }
\eeq
which is also determined by a holomorphic function, up to singularities
in the K\"{a}hler metric.
Positivity of the vacuum energy and the smooth behavior of (\ref{pot})
are then enough to preclude phase transitions -- although 
several phases can co-exist as degenerate vacua, the system
will never make a discontinuous jump from one phase to another.
We are merely making an analogous argument regarding 
$F$-component vacuum expectations which break SUSY 
dynamically. However, 
there is an important issue
to discuss here, namely that models with softly-broken SUSY {\it can} exhibit
phase transitions. As discussed in \cite{soft}, 
softly-broken models (such as QCD) can be obtained
from SUSY theories in the limit of large vacuum expectations for the 
$F$-components of spurion chiral
superfields, with the superpotential
remaining holomorphic in the spurion fields. 
We should try to understand what goes wrong with the usual 
arguments against phase transitions 
when the spurion $F$-components are turned on. In particular,
what suddenly allows the existence of phase transitions in these models?
Are there suddenly patches of singularities in the K\"{a}hler metric?
The answer is actually more mundane. When spurion fields with
non-zero $F$-components are present, terms in the K\"{a}hler potential can
easily mimic terms in the superpotential. Equation (\ref{Fvev}) no longer
applies, as there can be terms of the type
\beq
\label{terms}
F^*_{\rm spur} F_i ~~G( A_i^*, A_j )
\eeq
in the Lagrangian, where $G$ is some undetetermined function and $A_i$ are
the lowest components of the superfields $\Phi_i$. The effect of (\ref{terms}) on
(\ref{pot}) is to shift the potential by some non-holomorphic quantity. At 
low energies ($E << F^*_{\rm spur}$), 
these terms can dominate the superpotential contributions, and
can easily lead to phase transitions.

However, this loophole does not appear in our
analysis of spontaneous SUSY breaking. 
All non-zero $F$-components in models with 
spontaneous SUSY breaking must be proportional to the same SUSY
breaking scale -- there is no frozen spurion field $F_{\rm spur}$.
In that case
equation (\ref{Fvev}) always applies and a true singularity of the K\"{a}hler
metric is required for non-analytic behavior. The type of behavior we 
predict for models with spontaneous SUSY breaking is therefore
completely analogous to the absence of phase transitions in
models with unbroken SUSY.

A possible application of these observations is to the 
construction of new models
which break SUSY. Consider a model described by the bare Lagrangian
$\L_0$ which breaks SUSY. Now consider adding some new 
interaction\footnote{This can be, for example, the
gauging of some global symmetry already present in
the model. Then we can explore SUSY breaking as a function of the new
gauge coupling constant.}: $\L = \L_0 + g \O$. 
In many cases it is easy to investigate the effect of adding the
interaction $g \O$ when $g$ is infinitesimal. For example, often the
effect on the vacuum energy vanishes with $g$, and hence for sufficiently
small $g$ the vacuum energy will remain non-zero and SUSY will remain broken.
But because SUSY is broken in the region $g$ near zero, it will generically be
broken for all values of $g$ in the larger class of models. Interestingly, 
when $g$ is large the mechanism for SUSY breaking may differ significantly  from
the original mechanism in $\L_0$.  
(See the discussion below of the $3-2$ model
for an example of this.)
Alternatively, in some cases adding the extra interaction
$g \O$ will restore SUSY at non-zero $g$. In these cases 
the larger class of models
is seen as generically SUSY preserving, with $g=0$ an exceptional point of
SUSY breaking.

Another possiblity is the addition of new, massive particles to the original
model. In some cases it is possible to conclude that the addition of the
new particles does not change the status of SUSY breaking if they are 
sufficiently heavy. In this case, one can conclude that the status of SUSY
breaking remains the same as the mass of the heavy particles is
reduced, and they become relevant to the low-energy dynamics.

In section 3 we will discuss some examples of the above behavior,
 including some new models of SUSY breaking which are constructed in this way.

\section{Relation to the Witten index}

It is worthwhile to compare our conclusions to those that can be obtained
from the Witten index \cite{WI}. The Witten index is defined as
\beq
\rm Tr(-1)^F ~=~ n_0^B ~-~ n_0^F~~~,
\eeq
and counts the difference in the number of bosonic and fermionic zero
modes. It is a consequence of the SUSY algebra that  non-zero
energy states come in degenerate fermion-boson pairs, and hence continuous
deformations of a model cannot change $\rm{Tr(-1)^F}$, except in the
special case where a vacuum state comes in from or out to infinity 
along a pseudo-flat direction carrying a non-zero 
contribution to index.

When $\rm{Tr(-1)^F} \neq 0$ SUSY is clearly preserved, and the implications
are similar to what we concluded from holomorphy: generic variations of the
parameters leave SUSY intact, except for some special values where a vacuum
state escapes to infinity.
However, the case of
$\rm{Tr(-1)^F} = 0$ is ambiguous, because there may either be no zero energy
states (violating SUSY) or an even number of paired zero energy states 
(preserving SUSY). It is for vanishing Witten index that holomorphy provides some
interesting new information. As a parameter in the model is changed, it is possible 
for a pair of eigenstates with non-zero energy to flow 
down to zero energy, thereby restoring SUSY if it was initially broken. However,
what we learn from the holomorphy argument is that this pair of eigenstates cannot
``stick'' at zero energy: the pair must instead merely touch zero at some special
value of the parameter and then ``bounce'' back up to non-zero energy,
or alternatively only reach zero energy asymptotically 
as the parameter approaches infinity.
There is
nothing wrong with non-analytic behavior of the eigenvalues, 
as the energy of a state (\eg  see equation (\ref{pot}) ) is not
itself holomorphic in the parameters\footnote{This is in contrast to 
quantum systems with a {\it finite} number of degrees of freedom (\eg
  SUSY QM), where the groundstate energy is generally analytic
in the parameters.
Witten \cite{WI} pointed out that this implies that 
SUSY QM must behave in the manner advocated here for SUSY
field theories. However, his argument does not obviously extend 
to field theory
as in the limit of an infinite number of degrees of freedom analyticity
can fail. (This is related to the well-known loophole that allows for 
phase transitions and symmetry breaking in the infinite volume limit.)}. 
On the other hand, if 
$\rm{Tr(-1)^F} = 0$ but SUSY is preserved, we do not expect the zero energy
pair of eigenvalues to leave $E=0$ except at some special  point 
in parameter space,
where, \eg , the SUSY vacuum escapes to infinity. In the next section we will
see explicit examples of these types of behaviors.

\section{Examples}

Below we discuss some explicit examples of the general behavior described
in the previous sections. Included are some new models of SUSY
breaking constructed along previously described lines.

\subsection{Quantum deformation of moduli space: $\rm SU(2) ~+$ singlets}

SUSY breaking can arise from the quantum deformation of moduli space
\cite{ISS,IY,IT}. 
A simple example of a moduli space with a quantum deformed
constraint is $SU(2)$ with four doublet matter fields $Q_i$,
$i=1,\dots,4$.  The classical moduli space is parameterized by the
gauge invariants $M_{ij}= Q_i Q_j$ subject to the constraint $ {\rm
Pf}~M \equiv \epsilon^{ijkl} M_{ij} M_{kl} = 0$.  Quantum
mechanically, the constraint is modified to $ {\rm Pf}~M =
\Lambda_2^4$.  While the point $M_{ij}=0$ is part of the
classical moduli space, it does not lie on the quantum moduli space.

Supersymmetry is broken if the quantum modification of the  moduli
space is incompatible with a stationary superpotential, $W_{\phi}
\neq 0$. In the $SU(2)$ case, supersymmetry would be broken 
if there were $F$ terms
which only vanished for $M_{ij}=0$.  A simple realization of this \cite{IT}
is to add to the $SU(2)$ model given above, six singlet fields
$S^{ij}=-S^{ji}$, where $i,j=1,\dots,4$, with couplings 
\beq
W_{0} = \lambda ~ S^{ij} Q_i Q_j = \lambda~S^{ij} M_{ij}.  
\eeq

This
superpotential leaves invariant an $SU(4)_F$ flavor symmetry under
which $Q_i$ transform as $\bf{4}$, and $S^{ij}$ transform as $\bf{6}$.
There is an anomaly free $U(1)_R$ symmetry under which $R(Q)=0$
and $R(S)=2$.  Classically, there is a moduli space of supersymmetric vacua with
$M_{ij}=0$ and $S^{ij}$ arbitrary.  Quantum mechanically, the $S^{ij}$
equations of motion, $\lambda M_{ij}=0$, are incompatible with the
quantum constraint ${\rm Pf}~M=\Lambda_2^4$.  The classical moduli
space of supersymmetric vacua is completely lifted for $\lambda \neq
0$ as a result
of the quantum modification of the $SU(2)$ moduli space,
and supersymmetry is broken. 
 
Let us now consider some modifications of the theory. 
First, consider adding to the following interaction to the superpotential $W_0$:
\beq
W_1 ~=~ W_0 ~+~ g \epsilon_{ijkl} S^{ij} M^{kl}~.
\eeq
The new interaction, unlike $W_0$, is not invariant under $U(4)_F$
flavor rotations with determinant minus one. It thus violates a parity-like
symmetry of the original model.

The condition that $W_1$ be extremal with respect to variations in $S_{ij}$ 
now requires that
\beq
\lambda M_{ij} ~=~ -~g \epsilon_{ijkl} M^{kl}~.
\eeq
Combining this with the quantum constraint and extremality of $W_1$ with
respect to variations in $M_{ij}$ then requires
\beq
\label{spec}
\lambda^2 ~=~ 4 g^2~.
\eeq
We see that, for generic values of the new coupling g, SUSY remains
broken, with the exception of the special points satisfying (\ref{spec}).
From our previous discussion of the Witten index, we conclude that
at these special points a pair of boson/fermion states drops to zero
energy, maintaining $\rm Tr(-1)^F = 0$ but restoring SUSY.

It is interesting to note\footnote{We thank Y. Shirman for bringing this to our
attention.} that one could rewrite the model in terms of shifted 
singlet fields
\beq
\label{X}
X_{ij} ~=~\lambda S_{ij} ~+~g \epsilon_{ijkl} S^{kl}~.
\eeq 
Then the superpotential is simply $W_1 = X^{ij} M_{ij}$, and
minimizing with respect to $X_{ij}$ 
appears to break SUSY in the same manner as in
the original model with $W_0$. 
This is correct for generic
values of $\lambda$ and $g$, but for those satisfying (\ref{spec}) the above
conclusion fails as the transformation (\ref{X}) becomes non--invertible.
This leads to a singular 
K\"{a}hler metric in terms of $X_{ij}$, if one begins with a non-singular
K\"{a}hler metric in $S_{ij}$. This example illustrates how SUSY can
be restored at a special point due to a singularity in $g_{ij}$. It also
shows the danger of assuming that the K\"{a}hler metric remains smooth
while making arbitrary field redefinitions.

Alternatively, we can consider modifcations which explicitly
violate $U(1)_R$ symmetry,
enlarging the parameter space of the model.
The low energy effective superpotential will then
contain additional interactions, unconstrained by $U(1)_R$.
For a generic superpotential of this type, 
analysis of the extremality conditions
$W_{\phi} = 0$ shows that they can now be satisfied
when any of the $U(1)_R$--violating 
couplings are non-zero. Thus in this larger space of 
models SUSY is generically preserved, and the 
$U(1)_R$--preserving point is exceptional.
One can confirm that as the new couplings are taken to zero, 
the SUSY vacuum escapes to infinity.

Finally, we can consider the addition of additional massive
quark flavors. For instance, add an additional pair of $SU(2)$ doublets,
increasing the number of flavors to 3 (so $\rm N_f = N_c + 1$).
Then the $M_{ij}$ satisfy the quantum constraint
\beq
\label{constr}
({\rm Pf} ~M)~ \Bigl( {1 \over M} \Bigr)_{ij} ~=~ \Lambda^3_3~ m_{ij},
\eeq
where $\Lambda_3$ is related to the strong scale of the 3-flavor
model, and the indices now run from $1,...,6$. We will consider the 
case in which quarks $1-4$ remain massless and coupled to the
singlet fields, while $5,6$ are given the mass $m$.
Then the constraint (\ref{constr}) requires 
$M_{i5} =  M_{i6} = 0, ~i = 1,..6$, while
the light degrees of freedom satisfy a modified constraint
${\rm Pf}  M ~=~ \Lambda^3_3 m ~=~ \Lambda_2^4$.
Thus, the low energy dynamics is essentially that of the $\rm N_f = 2$
model, but with a modified strong coupling scale obtained by
matching at the heavy quark mass. SUSY remains broken for
all values of $m$, except at the special point $m = 0$, where again
a pair of boson/fermion states must drop to zero
energy.

\subsection{$\rm 3-2$ model; gauging global $\rm U(1)$; massive matter}

Another interesting example is provided by the 
3-2 model \cite{ADS,IT}. 
The matter content of the model is (here we indicate the
$SU(3) \times SU(2)$ charges): 
$P  ~(3,2),~       
L   ~(1,2), ~   
\overline{U} ~(\overline{3},1),~
\overline{D} ~(\overline{3},1).$  
This is just the one generation supersymmetric standard model
without hypercharge, the positron, or Higgs bosons.  Classically, this
model has a moduli space parameterized by three invariants: $Z=P^2
\overline U \overline D$, $X_1 = P L \overline D$, and $X_2 = P L
\overline U$.  There is another gauge invariant, $Y=P^3 L$, which
vanishes classically by Bose statistics of the underlying fields.  The
gauge group is completely broken for generic vacua on the classical
moduli space; the above invariants are the fields which are left
massless after the Higgs mechanism.  At the bare level there is a single
renormalizable coupling which can be added to the superpotential,
$
W_{0} = \lambda X_1.  
$
This superpotential leaves invariant
non-anomalous accidental 
$U(1)_R$ and $U(1)$ flavor symmetries, and 
completely lifts the classical moduli space.  Classically,
there is a supersymmetric ground state at the origin,
with the gauge symmetries unbroken. 

Non-perturbative gauge dynamics generate an additional term in the
effective superpotential; the exact effective superpotential
is fixed by holomorphy, symmetries, and an instanton
calculation to be
\beq
\label{super}
W={\Lambda _3^7\over Z}+
{\cal A}(Y- \Lambda_2^4) +
\lambda X_1.  
\eeq
where ${\cal A}$ is a Lagrange multiplier field. 
The first term is generated by instantons in the broken
$SU(3)$;
it is the usual dynamical superpotential familiar from
$SU(3)$ dynamics in the limit $\Lambda_3 >> \Lambda_2$.
The second term enforces the quantum deformed constraint 
$Y=P^3L=\Lambda_2^4$, which can be seen in the limit 
$\Lambda_2 >> \Lambda_3$. 
In this second limit we can neglect the $SU(3)$ dynamics and consider
the $SU(2)$ theory which has two flavors and a quantum deformed
constraint.
For nonsingular K\"{a}hler potential,
(\ref{super}) lifts the classical ground state
$Z=X_i=0$.  In the ground state of the quantum theory, both 
the $U(1)_R$ and supersymmetry 
are spontaneously broken.

There are several aspects of this
model which illustrate our previous discussion:

\noindent $\bullet$ We can vary the bare
coupling constants $\alpha_3, \alpha_2$.
SUSY is broken generically for all values, but via different mechanisms in the
two limits \cite{IT}. In the $\Lambda_2 >> \Lambda_3$ limit we have a 
quantum-deformed constraint, whereas in the opposite limit we have a 
dynamically generated superpotential.

\noindent $\bullet$ We can gauge the $U(1)$ flavor symmetry \cite{DNS}. 
SUSY is clearly still broken for
weak gauging, and must remain so even when the 
extra $U(1)$ is strongly coupled.

\noindent $\bullet$ Finally, we can add massive matter 
in vector like representations (section 6 of \cite{IT}).  
SUSY remains broken for generic values of the masses.

\subsection{$\rm SU(5)$ with $\bar{5} + 10$}

This model (along with $SO(10)$ with a single $\bf 16$) 
is one of the early candidate models for dynamical SUSY breaking
\cite{ADS},
but is difficult to study because it is strongly coupled and has no
flat directions.

Murayama  \cite{HM} proposed a technique to study such models, by the
introduction of extra vector like $5$ and $\bar{5}$ fields with mass $m$, 
which produce pseudo-flat directions. He 
finds SUSY breaking for non-zero values of $m$, and suggests that
this behavior must persist in the $m \rightarrow \infty$ limit,
using the vanishing Witten index as justification. However, 
Murayama's analysis still contains a potential loophole, as he originally
noted. Zero Witten index
does not preclude the energy of a pair of fermion/boson states from 
approaching zero at large $m$, restoring SUSY. 
(As happens in the models of section {\bf 3.1}
at the special points $\lambda^2 = 4 g^2$ or $m = 0$.)
Unfortunately, even
the holomorphy argument cannot quite close this loophole.
It tells us that as $m \rightarrow \infty$,
SUSY remains broken for generic $m$, but
cannot exclude the possibility that
SUSY breaking turns off asymptotically as $m \rightarrow \infty$.

\begin{flushleft} {\Large\bf Acknowledgments} \end{flushleft}

\noindent The authors are grateful to Nick Evans and Yuri Shirman 
for useful discussions and comments.
This work was supported in part under DOE contract DE-AC02-ERU3075.


\end{document}